\documentstyle[aps,prc]{revtex}
\begin{document}
\draft
\title{
Diffusion processes with Non-Markovian Transport Coefficients}
\author{Noboru Takigawa$^1$,
\thanks{E-mail address: takigawa@nucl.phys.tohoku.ac.jp}
Sakir Ayik$^2$, 
\thanks{E-mail address: ayik@tntech.edu}
and 
Sachie Kimura$^1$,
\thanks{E-mail address: sachie@nucl.phys.tohoku.ac.jp}
}
\address{$^1$ Department of Physics, Tohoku University, 
Sendai 980-8578, Japan \\
$^2$ Physics Department,
Tennessee Technological University, Cookeville, TN 38505
}
\date{\today}

\maketitle

\begin{abstract}

Employing time-dependent projection formalism, a  Fokker-Planck
equation with non-Markovian transport coefficients is derived for 
large amplitude collective motion. Properties of transport coefficients
for diffusion processes in a potential well and along a potential 
barrier are discussed, and their connection with the 
fluctuation-dissipation theorem is investigated. In the case of
diffusion in a potential well, the formalism naturally leads to
the generalization of the well known Einstein relation by including
quantum fluctuations in addition to thermal fluctuations.
Furthermore, it is shown that at low temperatures diffusion along 
potential barrier is significantly modified by quantum fluctuations.
Explicit expressions for transport coefficients are presented
for the case of the Caldeira-Leggett model.

\end{abstract}

\section{Introduction}

The fluctuation-dissipation theorem is a fundamental relationship 
in open system problems such as the Brownian motion \cite{weiss}. 
The most famous one is 
the Einstein relation, which connects the diffusion coefficient 
to the friction coefficient with the temperature 
as the proportional coefficient. 
This relationship can be easily derived if the 
bare potential for the Brownian particle, or more generally 
the collective motion, is flat and if the environment can be 
approximated by a canonical distribution. It has to be modified if 
the quantum fluctuation becomes comparable to the thermal fluctuation. 
When the bare potential for the collective motion is a parabolic well, 
the modified formula is well known. It resembles the 
original Einstein relation, 
in which temperature is
replaced by an effective temperature 
\cite{weiss,llsp}. The effective temperature
converges to the genuine temperature when the thermal energy dominates
the quantal zero point motion energy of the collective motion, and it 
converges to the zero point motion energy in the opposite limit.

However, there exist only a few investigations when the bare 
potential of the collective motion is not a potential well but a 
potential barrier \cite{ar94,hh00}. 
The aim of this paper is to address to this question. 
Following the 
standard treatment for deriving the fluctuation-dissipation relation
for motion in a potental well, we investigate the connection between
moments of the response and the correlation functions for the motion
along a potential barrier. 
We wish to emphasize that this study is important in various problems.
One of the examples is the stability of nuclear collective excitation 
\cite{ar94,aygs98}.
Another example is the fusion between 
two very heavy nuclei, which is used to try to 
synthesize superheavy elements. 
In this case, the fission barrier locates inside 
the fusion barrier. Consequently, 
it is not sufficient to overcome the fusion barrier in order 
for the two nuclei to fuse. They have to overcome also the fission barrier. 
One possible theoretical approach to this problem is to 
apply the Fokker Planck equation, which has been originally 
put forward by Kramers \cite{Kra40} to describe nuclear fission and chemical 
reactions, to describing the time evolution from the fusion point to 
the region inside the fission barrier. Though several pioneering works 
in this method have already been published \cite{awo}, 
it is not obvious whether 
one can apply the original Fokker Planck equation of Kramers 
as it is. For example, 
it is required to synthesize superheavy elements at temperatures as low as 
possible in order to sustain 
 the shell correction energy which yields the barrier against 
 fission by quantum effect. 
On the other hand, the curvature of the fission barrier is of the order of 
1 MeV.  One will then need to care about the quantum 
fluctuation. 
In this connection, 
it is highly desirable to generalize the original Fokker 
Planck equation by Kramers by taking quantum fluctuation 
into accont for the diffusion process along a potential barrier. 
The formalism presented here provides a first step in this direction.
Following a weak-coupling treatment, we derive a  generalized 
Fokker-Planck equation which incorporates quantum statistical effects
in terms of non-Markovian transport coefficients.

The paper is organized as follows.
In section II, we sketch the basic ideas of our theory to describe 
the time evolution of coupled collective and microscopic motions. 
It is a generalization of the quasi linear response theory in ref.
\cite{tak81} to include a quantum effect of the collective motion. 
We thus derive the generalized Fokker-Planck equation 
for the collective motion with non-Markovian transport coefficients. 
In section III, we discuss the properties of the force exerted 
by the microscopic motion, which we sometimes call the bath space, 
to determine the average classical trajectory of the collective motion. 
In section IV,  we represent the response and the correlation 
functions in terms of the spectral functions \cite{kubo}, 
and discuss their symmetry properties.
In section V, we discuss the memory times of the response and the correlation 
functions. In order to have some concrete ideas, we resort to the 
Caldeira-Leggett model \cite{cl} to this end. 
Section VI deals with the collective motion in a flat potential or 
in a parabolic well and derives the  well known fluctuation dissipation 
relation including the effect of quantum fluctuation. 
In sect.VII we consider the case of Ohmic dissipation 
of the Caldeira-Leggett model, and 
give the explicit expressions of the odd moment of the response function and 
the even moment of the correlation function for the motion in a 
potential well and the odd moment of the response function for the 
motion along a potential barrier. 
In sect.VIII we study the properties of the even moment 
of the correlation function for the collective 
motion along a parabolic potential 
barrier by taking the case of the Ohmic dissipation of the 
Caldeira-Leggett model with a cut-off function as an example.
We show the strong time dependence of the even moment of the correlation 
function and show that the asymptotic value as the time goes to infinity 
converges to a classical expectation value 
at high temperatures, but significantly deviates from it at low temperatures. 
We summarize the paper in sect. IX. 
Appendix A gives the main steps of the perturbation treatment 
of the coupled von-Neumann equations for the collective and microscopic 
motions and show how the reduced von-Neumann equation with 
non-Markovian transport coefficients is obtained.  
Appendix B is added to briefly sketch the symmetry property of the spectral 
function which is useful to derive the fluctuation-dissipation theorem. 
Appendix C collects some basic formulae and the spectral 
function of the Caldeira-Legget model. 
In Appendix D we present a general expression of the m-th moment 
of the response function for the Caldeira-Leggett model. 
We give an explicit expression of the linear friction for 
the case of Ohmic dissipation. Also we show that the linear 
friction is absent in the case of the super-Ohmic dissipation, and that 
the leading order is the n-th oder friction if the 
spectral density is proportional to $\omega^n$.   

\section{Basic equations of the quasi linear response theory}

\subsection{Galilei transformation}

We call the subspace of the collective motion and that of 
microscopic motions or environments as the subspaces A and B, respectively, 
and use A and B as the indices 
to distinguish the quantities referring to each subspace. The mass, 
coordinate and momentum for the collective motion are denoted by 
$M$, $q$ and $p$, respectively, while those for the environments by 
${m_i,x_i,p_i},i$ being 1,2..... We often use $x$ to denote the 
ensemble of the coordinates of the environmental degrees of freedom. 
As often done, quantum operators are distinguished from 
the corresponding c-numbers ${\cal O}(t)$ with ${\hat {\cal O}}$. 

We assume that the Hamiltonian for the total system is given by
\begin{eqnarray}
{\hat H}={\hat H}_A+{\hat H}_B+{\hat V}_c({\hat q},{\hat x})
\label{hamilt}
\end{eqnarray}
with
\begin{eqnarray}
{\hat H}_A= \frac{{\hat p}^2}{2M}+{\hat U}({\hat q})
\label{hamila}
\end{eqnarray}
where ${\hat U}({\hat q})$ and ${\hat V}_c({\hat q},{\hat x})$ are 
the bare potential for the collective motion 
and the coupling Hamiltonian, respectively. 
The time evolution of the density operator of the total sysytem
is determined by the von-Neumann equation,
\begin{eqnarray}
i\hbar \frac{\partial}{\partial t}{\hat \rho}(t)=[{\hat H},{\hat \rho}(t)]
\label{von-Neu}
\end{eqnarray}

We first introduce the classical variables $q(t)$ and $p(t)$ in the 
spirit of the Ehrenfest theorem by 
\begin{eqnarray}
q(t)=Tr({\hat q}{\hat \rho}(t)), \hskip 2.0cm 
p(t)=Tr({\hat p}{\hat \rho}(t)). 
\label{clavar}
\end{eqnarray}
Eq.\ref{von-Neu} leads to 
\begin{eqnarray}
{\dot q}(t)&=&\frac{p(t)}{M},
\label{velo}  \\ 
{\dot p}(t)&=&\frac{1}{i\hbar}Tr([{\hat p},{\hat H}]{\hat \rho}(t))
\nonumber \\
&=&\frac{1}{i\hbar} Tr(\{[{\hat p},{\hat U}({\hat q})]
+[{\hat p},{\hat V}({\hat q},{\hat x})]\}
{\hat \rho}(t))
\label{newton0}
\end{eqnarray}
We then introduce the Galilei transformation operator and the 
density operator in the Galilei transformed coordinate system by
\begin{eqnarray}
&&{\hat G}(t)=exp[\frac{i}{\hbar}(p(t){\hat q}-q(t){\hat p})]
\label{galilei}
\\
&&{\hat \rho}(t)={\hat G}(t){\hat \rho}_G(t){\hat G}^\dagger(t)
\label{gdensity}
\end{eqnarray}
It is easy to prove that
\begin{eqnarray}
i\hbar \frac{\partial}{\partial t}{\hat \rho}_G(t)=
[{\hat H}_G(t),{\hat \rho}_G(t)]
\label{gvon-Neu}
\end{eqnarray}
with
\begin{eqnarray}
{\hat H}_G(t)={\hat G}^\dagger(t){\hat H}{\hat G}(t)
-i\hbar {\hat G}^\dagger(t)\frac{\partial}{\partial t}{\hat G}(t)
\label{ghamil}
\end{eqnarray}

The Galilei transformation is convenient to describe a large amplitude 
collective motion, because one then needs to handle only the fluctuations 
around the mean values. Noting that,
\begin{eqnarray}
{\hat G}^\dagger(t){\hat q}{\hat G}(t)={\hat q}+q(t)
\\
{\hat G}^\dagger(t){\hat p}{\hat G}(t)={\hat p}+p(t)
\label{gtrans}
\end{eqnarray}
we expand ${\hat H}_G$ in the power series of the operators of the 
collective motion
\begin{eqnarray}
{\hat H}_G&=&c-number+
{\hat p}\{\frac{p(t)}{M}-{\dot q}(t) \}+
{\hat q}\{{\dot p}(t)+\frac{\partial U(q(t))}{\partial q(t)}+
\frac{\partial V_c(q(t),{\hat x})}{\partial q(t)}\}
\nonumber \\
&+&{\hat H}_B({\hat x})+V_c(q(t),{\hat x})+
\frac{{\hat p}^2}{2M}+\frac{1}{2}(\frac{\partial^2U}{\partial q^2}
+\frac{\partial^2V_c}{\partial q^2})_{q(t)}{\hat q}^2+{\hat {\cal V}}^{(3)}
\label{ghamilt}
\end{eqnarray}
where $(\frac{\partial^2U}{\partial q^2}+
\frac{\partial^2V_c}{\partial q^2})_{q(t)}$ is the value of 
$(\frac{\partial^2U}{\partial q^2}+\frac{\partial^2V_c}{\partial q^2})$ 
at $q=q(t)$ 
and ${\hat {\cal V}}^{(3)}$ 
is the third and higher order terms in fluctuating quantities.
We note that by taking the time derivative of the equation
\begin{eqnarray}
Tr({\hat p}{\hat \rho}_G(t))=0
\label{paverage}
\end{eqnarray}
we obtain
\begin{eqnarray}
{\dot p}(t)=-\frac{\partial U}{\partial q}-Tr(\frac{\partial V_c(q(t),{\hat x})}{\partial q}{\hat \rho}_G(t))+{\cal F}
\label{claeq}
\end{eqnarray}
where ${\cal F}$ 
is the force exerted by the third or higher order fluctuations 
and it reads
\begin{eqnarray}
{\cal F}=\frac{1}{3!}Tr\{(\frac{\partial^3U}{\partial q^3}
+\frac{\partial^3V_c}{\partial q^3}){\hat q}^3 {\hat \rho}_G(t) \}+........
\label{fforce}
\end{eqnarray}

We ignore the ${\hat {\cal V}}^{(3)}$ in eq.(\ref{ghamilt}) and 
correspondingly the force ${\cal F}$ in eq.(\ref{claeq}) in 
this work. On the other hand,  
the ${\hat q}^2$ term on the right hand side of eq.(\ref{ghamilt}), 
which has been ignored in 
\cite{tak81}, is essential to discuss the quantum 
fluctuation, which is the major issue in this paper. 
Using eqs.(\ref{velo}) and (\ref{claeq})
the Hamiltonian in the Galilei transformed coordinate system 
can be written as
\begin{eqnarray}
{\hat H}_G&=&c-number+
{\hat q}\cdot {\hat F}(q(t),{\hat x})
+{\hat h}_B(t)+
\frac{{\hat p}^2}{2M}+\frac{1}{2}C(q(t)){\hat q}^2
\label{ghamiltb}
\end{eqnarray}
where ${\hat h}_B(t)$, 
the effective Hamiltonian for the subspace B at time t, 
the effective coupling force ${\hat F}(q(t),{\hat x})$ 
and the curvature parameter $C(q(t))$ are 
given by
\begin{eqnarray}
& &{\hat h}_B(t)={\hat H}_B({\hat x})+V_c(q(t),{\hat x})
\label{ehamilB} \\
& &{\hat F}(q(t),{\hat x})=
\frac{\partial V_c(q(t),{\hat x})}{\partial q}-
Tr(\frac{\partial V_c(q(t),{\hat x})}{\partial q}{\hat \rho}_G(t))
\label{ecforce} \\
& &C(q(t))=(\frac{\partial^2U}{\partial q^2}
+\frac{\partial^2V_c}{\partial q^2})_{q(t)}
\label{curva}
\end{eqnarray}
The von-Neumann equation for ${\hat \rho}_G(t)$ reads
\begin{eqnarray}
i\hbar \frac{\partial}{\partial t}{\hat \rho}_G(t)=
[{\hat h}_B(t)+{\hat q}\cdot {\hat F}(q(t),{\hat x})
+
\frac{{\hat p}^2}{2M}+\frac{1}{2}C(q(t)){\hat q}^2, {\hat \rho}_G(t)].
\label{gvonNeu}
\end{eqnarray}

\subsection{Fokker-Planck equation for the collective motion}

We now introduce the time evolution operator for the subspace B, 
${\hat u}_B$, by
\begin{eqnarray}
i\hbar \frac{\partial}{\partial t}{\hat u}_B(t,t_0)={\hat h}_B(t)
{\hat u}_B(t,t_0)
\label{evolution}
\end{eqnarray}
with the initial condition
\begin{eqnarray}
{\hat u}_B(t_0,t_0)=1
\label{evolutionb}
\end{eqnarray}
and the density operator ${\hat D}(t)$ by
\begin{eqnarray}
{\hat \rho}_G(t)={\hat u}_B(t,t_0){\hat D}(t){\hat u}_B^\dagger(t,t_0)
\label{ddensity}
\end{eqnarray}
The density operator ${\hat D}(t)$ obeys
\begin{eqnarray}
i\hbar \frac{\partial}{\partial t}{\hat D}(t)=
[{\hat q}\cdot {\hat f}(q(t),{\hat x}),{\hat D}(t)]+
[\frac{{\hat p}^2}{2M}+\frac{1}{2}C(q(t)){\hat q}^2, {\hat D}(t)]
\label{DvonNeu}
\end{eqnarray}
where
\begin{eqnarray}
{\hat f}(t)={\hat u}^\dagger_B(t,t_0){\hat F}{\hat u}_B(t,t_0)
\label{fforcel}
\end{eqnarray}

We introduce the reduced density operators for the subspaces A and B by
\begin{eqnarray}
{\hat D}_A(t)=Tr_B{\hat D}(t), \hskip 2cm {\hat D}_B(t)=Tr_A{\hat D}(t)
\label{dsub}
\end{eqnarray}
and the correlation density by
\begin{eqnarray}
{\hat D}_C(t)={\hat D}(t)-{\hat D}_A(t)\cdot {\hat D}_B(t)
\label{dcorre}
\end{eqnarray}
These density operators obey the following coupled equations
\begin{eqnarray}
i\hbar \frac{\partial}{\partial t}{\hat D}_A(t)&=&
Tr_B[{\hat q}\cdot {\hat f}(q(t),{\hat x}),{\hat D_C}(t)]+
[\frac{{\hat p}^2}{2M}+\frac{1}{2}C(q(t)){\hat q}^2, {\hat D}_A(t)]
\label{DAvonNeu} \\
i\hbar \frac{\partial}{\partial t}{\hat D}_B(t)&=&
Tr_A[{\hat q}\cdot {\hat f}(q(t),{\hat x}),{\hat D_C}(t)]
\label{DBvonNeu} \\
i\hbar \frac{\partial}{\partial t}{\hat D}_C(t)&=&
[{\hat q}\cdot {\hat f}(q(t),{\hat x}),{\hat D_A}(t)\cdot{\hat D_B}(t)]
+(1-{\hat D_A}(t)Tr_A-{\hat D_B}(t)Tr_B)
[{\hat q}\cdot {\hat f}(q(t),{\hat x}),{\hat D_C}(t)]
\nonumber \\
&+&[\frac{{\hat p}^2}{2M}+\frac{1}{2}C(q(t)){\hat q}^2, {\hat D}_C(t)]
\label{DCvonNeu}
\end{eqnarray}

Following a similar treatment presented in \cite{tak81}, whose major steps 
are briefly described in the Appendix A, 
we derive a diffusion equation for the collective motion by 
handling eqs.(\ref{DAvonNeu}) through (\ref{DCvonNeu}) 
in the second order perturbation theory with respect to the 
fluctuating force ${\hat f}$. 
In the derivation, we
ignore the time dependence of the curvature parameter C,  
and introduce the moment functions ${\cal C}(t,t_1)$ and 
${\cal S}(t,t_1)$ by 
\begin{eqnarray}
{\hat u}^{(0)}_A(t,t_1){\hat q}{\hat u}^{(0)\dagger}_A(t,t_1)=
{\cal C}(t,t_1){\hat q}-{\cal S}(t,t_1){\hat p}
\label{mfunc}
\end{eqnarray}
where
\begin{eqnarray}
{\hat u}^{(0)}_A(t,t_1)=exp[\frac{1}{i\hbar}\{
\frac{{\hat p}^2}{2M}+\frac{1}{2}C(q(t)){\hat q}^2\}(t-t_1)]
\label{evolA}
\end{eqnarray}
They are 
\begin{eqnarray}
{\cal C}(t,t_1)&=&cos[\Omega(t-t_1)] 
\label{mfuncCP} \\
{\cal S}(t,t_1)&=&\frac{1}{{\sqrt {MC}}}sin[\Omega(t-t_1)] 
\label{mfuncSP}
\end{eqnarray}
with 
\begin{eqnarray}
\Omega={\sqrt{\frac{C}{M}}}
\label{OmegaP}
\end{eqnarray}
when $C\geq0$ and 
\begin{eqnarray}
{\cal C}(t,t_1)&=&cosh[\Omega(t-t_1)] 
\label{mfuncCN} \\
{\cal S}(t,t_1)&=&\frac{1}{{\sqrt {M \vert C \vert}}}sinh[\Omega(t-t_1)] 
\label{mfuncSN}
\end{eqnarray}
with 
\begin{eqnarray}
\Omega={\sqrt{\frac{\vert C \vert}{M}}}
\label{OmegaN}
\end{eqnarray}
when $C<0$.
  
Furthermore we define the response and the correlation functions by
\begin{eqnarray}
\chi^{(-)}_{\alpha \beta}(t,t_1)&=&\frac{i}{\hbar}
Tr_B([{\hat f}_\alpha(t),{\hat f}_\beta(t_1)]{\hat D}_B(t_1))
\label{rfunct} \\
\chi^{(+)}_{\alpha \beta}(t,t_1)&=&\frac{1}{2}
Tr_B([{\hat f}_\alpha(t),{\hat f}_\beta(t_1)]_+{\hat D}_B(t_1))
\label{cfunct}
\end{eqnarray}
and their moments by
\begin{eqnarray}
\chi^{(-E)}_{\alpha \beta}(t)&=&\int^t_{t_0}dt_1 {\cal C}(t,t_1)
\chi^{(-)}_{\alpha \beta}(t,t_1)
\label{REmoment} \\
\chi^{(-O)}_{\alpha \beta}(t)&=&\int^t_{t_0}dt_1 {\cal S}(t,t_1)
\chi^{(-)}_{\alpha \beta}(t,t_1)
\label{ROmoment} \\
\chi^{(+E)}_{\alpha \beta}(t)&=&\int^t_{t_0}dt_1 {\cal C}(t,t_1)
\chi^{(+)}_{\alpha \beta}(t,t_1)
\label{CEmoment} \\
\chi^{(+O)}_{\alpha \beta}(t)&=&\int^t_{t_0}dt_1 {\cal S}(t,t_1)
\chi^{(+)}_{\alpha \beta}(t,t_1)
\label{COmoment} 
\end{eqnarray}
In these equations the lower indices $\alpha $ and $\beta $ represent the 
spatial component in the case when the diffusion process takes place in a 
multi-dimensional space. 
The upper indices $E$ and $O$ stand for even and odd, respectively. 
The von-Neumann equation for ${\hat D}_A(t)$ is then given by
\begin{eqnarray}
\frac{\partial}{\partial t}{\hat D}_A(t)&=&
\frac{1}{i\hbar}[\frac{{\hat p}^2}{2M}+\frac{1}{2}C{\hat q}^2,{\hat D}_A(t)]
\nonumber \\
&-&\frac{1}{\hbar^2}[{\hat q}_\alpha,
[\chi^{(+E)}_{\alpha \beta}(t){\hat q}_\beta-
\chi^{(+O)}_{\alpha \beta}(t){\hat p}_\beta,{\hat D}_A(t)]]
\nonumber \\
&-&\frac{1}{2i\hbar}[{\hat q}_\alpha,
[\chi^{(-E)}_{\alpha \beta}(t){\hat q}_\beta-
\chi^{(-O)}_{\alpha \beta}(t){\hat p}_\beta,{\hat D}_A(t)]_+]
\nonumber \\
\label{DAvonNeuF} 
\end{eqnarray}
Taking the Wigner transform of each side of eq.(\ref{DAvonNeuF})
we obtain the Fokker Planck equation for  
the Wigner distribution function for the subspace 
of the collective motion $D_{AW}(q,p,t)$ \cite{tak81}
defined by
\begin{eqnarray}
D_{AW}(p,q,t)=\int^{\infty}_{-\infty}dr e^{-ipr/\hbar}
<q+\frac{1}{2}r \vert {\hat D}_A(t) \vert q-\frac{1}{2}r>
\label{WDF} 
\end{eqnarray}
The result reads
\begin{eqnarray}
\frac{\partial}{\partial t} D_{AW}(t)
&=&(-\frac{1}{M}p_\alpha \frac{\partial }{\partial q_\alpha}
+Cq_\alpha\frac{\partial}{\partial p_\alpha}
-\chi^{(-E)}_{\alpha \beta} q_\beta \frac{\partial }{\partial p_\alpha}
+\frac{1}{M}\chi^{(-O)}_{\alpha \beta}
\frac{\partial }{\partial p_\alpha} p_\beta)D_{AW}(p,q,t)
\nonumber \\
&+&(\frac{1}{M}\chi^{(+O)}_{\alpha \beta} \frac{\partial^2}{\partial p_\alpha 
\partial q_\beta}
+\chi^{(+E)}_{\alpha \beta}
\frac{\partial^2}{\partial p_\alpha \partial p_\beta}
)D_{AW}(p,q,t)
\label{FPEQ} 
\end{eqnarray}
This is a generalization of the well known Fokker Planck 
equation in literatures \cite{Kra40,Bec78}. Although it appears to be
a Markovian equation, the memory effects are effectively
incorporated in the transport coefficients. As a result, the 
formalism provides a basis for describing quantum statistical
effects of collective motion. We also note that, since the derivation
of eq.(\ref{FPEQ}) 
is based on a perturbation theory with respect to the fluctuating force
${\hat f}$, the formalism is valid 
in the weak coupling limit, 
where the time scales of the response and correlation 
functions, i.e. their memory times $\tau^{(-)}_c$ and $\tau^{(+)}_c$ , 
are much shorter than the 
time scale, i.e. the relaxation time $\tau_R$, of the collective 
motion 
\cite{van Hove55,acw}. 
In sect.V, we discuss the properties of the response and 
correlation functions for the 
Caldeira-Leggett model, and show that their memory times 
$\tau^{(-)}_c$ and $\tau^{(+)}_c$ get very small if the cut-off 
frequency is large. This is consistent with the well known fact that 
the memory time gets very short if the collective motion couples with 
many incoherent degrees of freedom.

\subsection{Coupled equations to determine fluctuations of the 
collective motion}

We note that the average values of $q$ and $p$ are zero in the Galilei 
transformed space, and that the solution of eq.(\ref{FPEQ}) is a 
Gaussian. Therefore, one can put
\begin{eqnarray}
D_{AW}(p,q,t)&=&
\frac{1}{(2\pi)^{3/2}\Delta ^{1/2}}
\times 
exp\left[-\frac{1}{2\Delta}\sum_{i,j=1,2}y_iy_j{\tilde \sigma}_{i,j}
\right]
\label{wdfd} \\
\Delta&=&\sigma_{qq}\sigma_{pp}-\sigma_{qp}^2
\label{msflu} 
\end{eqnarray}
where $y_1=q$ and $y_2=p$ and the 2$\times $2 matrix ${\tilde \sigma}$ is 
the inverse matrix of the 2$\times $2 matrix 
\begin{eqnarray}
\left(
\begin{array}{ll}
\sigma_{qq} & \sigma_{qp} \\
\sigma_{qp} & \sigma_{pp}
\end{array}
\right)
\end{eqnarray}
which determines the fluctuations, i.e. the mean square deviations from the 
average values. 
The values of $\sigma_{ij}$ are obtained by solving the 
following coupled equations given by eq.(\ref{FPEQ}),
\begin{eqnarray}
\frac{d}{dt}
\left(
\begin{array}{l}
\sigma_{qq}(t) \\
\sigma_{qp}(t) \\
\sigma_{pp}(t) 
\end{array}
\right)
=
\left(
\begin{array}{lll}
0 & \frac{2}{M} & 0 \\
-(C-\chi^{(-E)}) & -\frac{\chi^{(-O)}}{M} & \frac{1}{M} \\
0 & -2(C-\chi^{(-E)}) & -2\frac{\chi^{(-O)}}{M}
\end{array}
\right)
\left(
\begin{array}{l}
\sigma_{qq}(t) \\
\sigma_{qp}(t) \\
\sigma_{pp}(t) 
\end{array}
\right)
+
\left(
\begin{array}{l}
0 \\
\frac{\chi^{(+O)}}{M} \\
2\chi^{(+E)} 
\end{array}
\right)
\label{ceqflu} 
\end{eqnarray}

\subsection{The Wigner distribution function 
in the original space}

The Wigner distribution function for the 
collective motion in the original space fixed frame 
is given by
\begin{eqnarray}
\rho_{AW}(q,p,t)=D_{AW}(q-q(t),p-p(t),t)
\label{wdfos}
\end{eqnarray}
once the Wigner distribution function in the Galilei transformed space 
$D_{AW}$ is obtained from eqs.(\ref{wdfd}) and (\ref{ceqflu}).

\section{Classical trajectory of the collective motion}

The classical trajectory of the collective motion given by $q(t)$ and $p(t)$ 
are determined by eqs.(\ref{velo}) and (\ref{claeq}).
Let us express the force induced by the coupling to microscopic 
motions in terms of the response function.
It is useful, for example, to discuss the fluctuation-dissipation theorem. 
We first note that to the same order as eq.(\ref{DAvonNeuF}) 
the density operator for the subspace B,
${\hat \rho}_B(t)=Tr_A{\hat \rho}(t)$, 
obeys the following von-Neumann equation
\begin{eqnarray}
i\hbar \frac{\partial}{\partial t}{\hat \rho}_B(t)=
[{\hat h}_B(t),{\hat \rho}_B(t)]
\label{von-NeuB}
\end{eqnarray}
We determine ${\hat \rho}_B(t)$ by a perturbation theory up to the second 
order of the coupling $V_c$. The effective force exerted on the
collective motion from the bath space is then given by 
\begin{eqnarray}
&F&_{BA}=-Tr_B(\frac{\partial V_c(q(t),{\hat x})}{\partial q}{\hat \rho}_B(t))
\label{FBA1} \\
&=&-Tr_B(\frac{\partial V_c(q(t),{\hat x})}{\partial q}
e^{\frac{1}{i\hbar}{\hat H}_B(t-t_0)}{\hat \rho}_B(t_0)
e^{-\frac{1}{i\hbar}{\hat H}_B(t-t_0)})
\nonumber \\
&-&\frac{1}{i\hbar}\int^t_{t_0}dt_1 
Tr_B(\frac{\partial V_c(q(t),{\hat x})}{\partial q}
[e^{\frac{1}{i\hbar}{\hat H}_B(t-t_1)}V_c(q(t_1),{\hat x})
e^{-\frac{1}{i\hbar}{\hat H}_B(t-t_1)},
e^{\frac{1}{i\hbar}{\hat H}_B(t-t_0)}{\hat \rho}_B(t_0)
e^{-\frac{1}{i\hbar}{\hat H}_B(t-t_0)}])
\label{FBA2} 
\end{eqnarray}

\subsection{Limit of ignoring the quantal effect}

There are two alternative ways to write down the 
connection between the friction tensor and the moments of the 
response function. The standard way is to ignore the quantum or 
the memory effect and relate the friction coefficient to the 
first moment of the response function. 
As often done, in this approach, 
we approximate ${\hat \rho}_B(t_0)$ by a canonical distribution 
${\hat \rho}^{(eq)}_B(t)=e^{\beta(t) ({\cal F}-{\hat H}_B)}$, 
ignore the time variation of $\beta$, 
and expand $V_c(q(t_1),{\hat x})$ in the square bracket on the r.h.s. of 
eq.(\ref{FBA2}) around $t_1=t$. We then obtain 
\begin{eqnarray}
F_{BA}&=&-Tr_B(\frac{\partial V_c(q(t),{\hat x})}{\partial q}
{\hat \rho}^{(eq)}_B(t))
\nonumber \\
&-&\frac{1}{i\hbar}\int^{t-t_0}_0 d\tau 
Tr_B([\frac{\partial V_c(q(t),{\hat x})}{\partial q},
e^{\frac{1}{i\hbar}{\hat H}_B\tau} V_c(q(t),{\hat x})
e^{-\frac{1}{i\hbar}{\hat H}_B\tau}]{\hat \rho}^{(eq)}_B(t))
\nonumber \\
&-&{\dot q}(t) \chi^{(-1)}(t)+.....
\label{friction}
\end{eqnarray}
where the first moment of the response function 
$\chi^{(-1)}(t)$ is defined by
\begin{eqnarray}
\chi^{(-1)}_{\alpha \beta}(t)=\int^t_{t_0}dt_1 (t-t_1)
\chi^{(-)}_{\alpha \beta}(t,t_1)
\label{R1moment} 
\end{eqnarray}
Eq.(\ref{friction}) is the expansion with respect to the moments of 
the response function, whose higher order terms can be ignored if 
the memeory time is short enough. 
It shows that the linear friction coefficient is given by the 
first moment of the response function in such cases. 
Note, however, that it becomes zero 
in some cases. As we show in Appendix D, the super-Ohmic dissipation 
is one of such examples.

\subsection{Inclusion of quantum effect}

The second way is to 
take the quantum or the memory effects into account and 
represent the friction coefficient in terms of the odd moment 
of the response function $\chi^{(-O)}$ to be consistent with eq.(\ref{FPEQ}). 
In this method, 
we first remark the following relationship between the classical variables 
at two different times,
\begin{eqnarray}
q(t_1)={\cal C}(t,t_1)q(t)-M{\cal S}(t,t_1){\dot q}(t).
\label{tevoho} 
\end{eqnarray}
This is valid when the classical motion evolves along a parabolic 
potential well or a parabolic barrier without friction
,i.e. in the harmonic approximation. We expand 
$V_c(q(t_1),{\hat x})$ in the last term on the r.h.s. of eq.(\ref{FBA2})
in powers of $q(t_1)-q(t)$ around $q(t)$, and use eq.(\ref{tevoho}).
This is allowed to the lowest order of the coupling strength. 
We then obtain 
\begin{eqnarray}
F_{BA}&=&-Tr_B(\frac{\partial V_c(q(t),{\hat x})}{\partial q}
{\hat \rho}^{(eq)}_B(t))
\nonumber \\
&-&\frac{1}{i\hbar}\int^{t-t_0}_0 d\tau 
Tr_B([\frac{\partial V_c(q(t),{\hat x})}{\partial q},
e^{\frac{1}{i\hbar}{\hat H}_B\tau} V_c(q(t),{\hat x})
e^{-\frac{1}{i\hbar}{\hat H}_B\tau}]{\hat \rho}^{(eq)}_B(t))
\nonumber \\
&+& q(t) \int^t_{t_0} dt_1 \left[{\cal C}(t,t_1)-1 \right] \chi^{(-)}(t,t_1)
\nonumber \\
&-&M{\dot q}(t) \chi^{(-O)}(t)+.....
\label{frictionB}
\end{eqnarray}

We note that the odd moment of the response function 
$\chi^{(-O)}_{\alpha \beta}(t)$ converges to the first moment 
of the response function 
in the limit of small $\Omega $, i.e. in the limit of small $C$. 
More precisely, for small $\Omega$
\begin{eqnarray}
\chi^{(-O)}(t)
=\left\{
\begin{array}{ll}
\frac{1}{M}\chi^{(-1)}(t)-{\cal O}(\Omega^2) & (C > 0) \\
\frac{1}{M}\chi^{(-1)}(t)+{\cal O}(\Omega^2) & (C < 0) 
\end{array}\right.
\label{RES1-O} 
\end{eqnarray}
The results in eqs.(\ref{friction}) and (\ref{frictionB}) 
agree to each other in this limit.

\section{Properties of the response and the correlation functions 
in the linear response theory}

\subsection{Spectral function}

We first note that $Tr_B({\hat f}_\alpha(t){\hat f}_\beta(t_1)
{\hat D}_B(t_1))$, which appears in the definition of 
the response and the correlation 
functions, can be rewritten as
\begin{eqnarray}
Tr_B({\hat f}_\alpha(t){\hat f}_\beta(t_1){\hat D}_B(t_1))
=Tr_B({\hat F}_\alpha(t){\hat u}_B(t,t_1){\hat F}_\beta(t_1)
{\hat u}^\dagger_B(t,t_1){\hat \rho}_B(t))
\label{comp1} 
\end{eqnarray}
where
\begin{eqnarray}
{\hat u}_B(t,t_1)={\hat u}_B(t,t_0){\hat u}^\dagger_B(t_1,t_0)
\label{tevolm} 
\end{eqnarray}
Similarly,
\begin{eqnarray}
Tr_B({\hat f}_\beta(t_1){\hat f}_\alpha(t){\hat D}_B(t_1))
=Tr_B({\hat u}_B(t,t_1){\hat F}_\beta(t_1){\hat u}^\dagger_B(t,t_1)
{\hat F}_\alpha(t){\hat \rho}_B(t))
\label{comp2} 
\end{eqnarray}

We consider the collective motion in the weak 
coupling limit, for which $\tau^{(-,+)}_c<<\tau_R$, so that 
the change of the collective variable during the 
decay time $\tau^{(-,+)}_c$ of the collision kernel is small. 
Then, we can approximate these relations by 
\begin{eqnarray}
Tr_B({\hat f}_\alpha(t){\hat f}_\beta(t_1){\hat D}_B(t_1))
&\approx&Tr_B({\hat F}_\alpha(t){\hat u}_B(t,t_1){\hat F}_\beta(t)
{\hat u}^\dagger_B(t,t_1){\hat \rho}_B(t)) 
\label{acomp1} \\
Tr_B({\hat f}_\beta(t_1){\hat f}_\alpha(t){\hat D}_B(t_1))
&\approx&Tr_B({\hat u}_B(t,t_1){\hat F}_\beta(t){\hat u}^\dagger_B(t,t_1)
{\hat F}_\alpha(t){\hat \rho}_B(t))
\label{acomp2} 
\end{eqnarray}
where the time evolution operator ${\hat u}_B(t,t_1)$ is given by 
\begin{eqnarray}
{\hat u}_B(t,t_1)=Te^{\frac{1}{i\hbar}\int^t_{t_1}
{\hat h}_B(t')dt'}
\end{eqnarray}
with the time ordering operator $T$. 
For time intervals much shorter than the relaxation
time of the collective motion, it can be apparoximated as,
\begin{eqnarray}
{\hat u}_B(t,t_1)={\hat u}_B(t;\tau=t-t_1)
\approx e^{\frac{1}{i\hbar}{\hat h}_B(t)\tau}
\end{eqnarray}

We now define the spectral functions by
\begin{eqnarray}
J_{\alpha,\beta}(t;\omega)
&=&\int^{+\infty}_{-\infty}d\tau e^{i\omega \tau}
Tr_B({\hat F}_\alpha(t){\hat u}_B(t;\tau){\hat F}_\beta(t)
{\hat u}^\dagger_B(t;\tau){\hat \rho}_B(t))
\label{specfun} \\
&\approx&\int^{+\infty}_{-\infty}d\tau e^{i\omega \tau}
Tr_B({\hat F}_\alpha(t)e^{-\frac{i}{\hbar}{\hat h}_B(t)\tau}{\hat F}_\beta(t)
e^{\frac{i}{\hbar}{\hat h}_B(t)\tau}{\hat \rho}_B(t))
\label{aspecfun} 
\end{eqnarray}
Inversion of these equations yields,
\begin{eqnarray}
Tr_B({\hat F}_\alpha(t){\hat u}_B(t;\tau){\hat F}_\beta(t)
{\hat u}^\dagger_B(t;\tau){\hat \rho}_B(t))
&=&\int^{+\infty}_{-\infty}\frac{d\omega}{2\pi}e^{-i\omega \tau}
J_{\alpha,\beta}(t;\omega)
\label{ispecfuna} \\
Tr_B({\hat u}_B(t;\tau){\hat F}_\beta(t_1){\hat u}^\dagger_B(t;\tau)
{\hat F}_\alpha(t){\hat \rho}_B(t))
&=&\int^{+\infty}_{-\infty}\frac{d\omega}{2\pi}e^{i\omega \tau}
J_{\beta,\alpha}(t;\omega)
\label{ispecfunb} 
\end{eqnarray}

\subsection{Symmetry properties}

We introduce a further  approximation 
by replacing the density operator of the subspace B 
with a canonical distribution,
\begin{eqnarray}
{\hat \rho}_B(t)\approx exp\{\beta(t)({\cal F}-{\hat h}_B(t))\}
\label{cano} 
\end{eqnarray}
As we show in Appendix B, the resultant specral function 
has the following symmetry property
\begin{eqnarray}
J_{\beta,\alpha}(t;\omega)=e^{\beta(t)\hbar \omega}J_{\alpha,\beta}(t;-\omega)
\label{symspec} 
\end{eqnarray}
One can then rewrite eq.(\ref{ispecfunb}) as
\begin{eqnarray}
Tr_B({\hat u}_B(t;\tau){\hat F}_\beta(t_1){\hat u}^\dagger_B(t;\tau)
{\hat F}_\alpha(t){\hat \rho}_B(t))
=\int^{+\infty}_{-\infty}\frac{d\omega}{2\pi}e^{-i\omega \tau}
e^{-\beta(t)\hbar \omega}J_{\alpha,\beta}(t;\omega)
\label{ispcfunbi} 
\end{eqnarray}
Consequently, the response and the correlation functions can be 
expressed as
\begin{eqnarray}
\chi^{(+)}_{\alpha,\beta}(t,t_1)=\chi^{(+)}_{\alpha,\beta}(t;\tau=t-t_1)
&=&\int^{\infty}_{-\infty}\frac{d\omega}{2\pi}
e^{-i\omega \tau}\frac{1+e^{-\beta(t)\hbar \omega}}{2}
J_{\alpha,\beta}(t;\omega)
\label{cspec1} \\
\chi^{(-)}_{\alpha,\beta}(t,t_1)=\chi^{(-)}_{\alpha,\beta}(t;\tau=t-t_1)
&=&\int^{\infty}_{-\infty}\frac{d\omega}{2\pi}
e^{-i\omega \tau}\frac{i}{\hbar}\{1-e^{-\beta(t)\hbar \omega}\}
J_{\alpha,\beta}(t;\omega)
\label{rspec1} 
\end{eqnarray}
Also, one can prove the following symmetry properties 
\begin{eqnarray}
\chi^{(+)}_{\alpha,\alpha}(t;-\tau)&=&\chi^{(+)}_{\alpha,\alpha}(t;\tau)
\label{csym} \\
\chi^{(-)}_{\alpha,\alpha}(t;-\tau)&=&-\chi^{(-)}_{\alpha,\alpha}(t;\tau)
\label{rsym} 
\end{eqnarray}
Furthermore, if we define
\begin{eqnarray}
{\tilde {\chi}}^{(+)}_{\alpha,\alpha}(t;\omega)&=&
\int^{\infty}_0 d\tau~ cos \omega \tau ~\chi^{(+)}_{\alpha,\alpha}(t;\tau)
\label{coschi} \\
{\tilde {\chi}}^{(-)}_{\alpha,\alpha}(t;\omega)&=&
\int^{\infty}_0 d\tau~ sin \omega \tau ~\chi^{(-)}_{\alpha,\alpha}(t;\tau)
\label{sinchi} 
\end{eqnarray}
then we obtain
\begin{eqnarray}
{\tilde {\chi}}^{(+)}_{\alpha,\alpha}(t;\omega)&=&
\frac{1}{2}\frac{1+e^{-\beta(t)\hbar \omega}}{2}
J_{\alpha,\alpha}(t;\omega)
\label{coscspec} \\
{\tilde {\chi}}^{(-)}_{\alpha,\alpha}(t;\omega)&=&
\frac{1}{2\hbar}\{1-e^{-\beta(t)\hbar \omega}\}
J_{\alpha,\alpha}(t;\omega)
\label{sinrspec} 
\end{eqnarray}

\section{Memory times of the response and correlation functions}

An important issue to further proceed is the memory times of 
the response and the correlation functions. 
In order to see their connection to 
various quantities of the system such as the properties of the 
environments, or the coupling Hamiltonian and the temperature, we consider 
explicitly the case of Feynman-Vernon model \cite{fv}, which is a 
popular model in open system problems and 
has been widely used to discuss the macroscopic quantum tunneling since the 
seminal work of Caldeira and Leggett \cite{cl}. Here we discuss the case 
of Ohmic dissipation. 
It is easy to show that the response function is given by
\begin{eqnarray}
\chi^{(-)}_{\alpha,\alpha}(t;\tau)
&=&\frac{2}{\pi} (\frac{df}{dq})^2 \eta
\int^{\infty}_0 ~sin(\omega \tau) \omega g(\omega)~~d\omega 
\label{clres} 
\end{eqnarray}
where $g(\omega)$ is the cut off function (see Appendix C). 
Using its explicit form, we have  
\begin{eqnarray}
\chi^{(-)}_{\alpha,\alpha}(t;\tau)
&=& 
\left\{
\begin{array}{ll}
\frac{1}{2{\sqrt \pi}} (\frac{df}{dq})^2 \eta
\cdot \omega_c^3 \cdot \tau \cdot e^{-\frac{1}{4} \omega^2_c 
\tau^2} & (Gaussian~~ cut-off) \\
(\frac{df}{dq})^2 \eta
\cdot \omega_c^2 \cdot e^{-\omega_c\tau} & (Drude~~regularization)
\end{array}\right.
\label{clresgd} 
\end{eqnarray} 
Eq.(\ref{clresgd}) shows that the 
memory time gets shorter for larger cut-off frequency. 
On the other hand, the correlation function is given by
\begin{eqnarray}
\chi^{(+)}_{\alpha,\alpha}(t;\tau)
&=&\frac{\hbar \eta}{\pi} (\frac{df}{dq})^2 \cdot T \cdot 
\int^{\infty}_{-\infty} d\omega ~cos(\omega \tau) 
\cdot \frac{\omega}{2T}coth(\frac{\beta(t)\hbar \omega}{2}) 
\cdot g(\omega) 
\label{clcor} 
\end{eqnarray}

Figure 1 shows the correlation function 
$\chi^{(+)}(t,\tau)$ as a function of $\tau $ for three 
values of the cut-off frequency and for a given temperature T=1 MeV. 
Since we consider a linear coupling, $\chi^{(+)}(t;\tau)$ is 
independent of time t, so that we denoted the ordinate as 
$\chi^{(+)}(\tau)$ instead of $\chi^{(+)}(t,\tau)$. 
The correlation function is normalized to 1.0 at 
$\tau=0$. It decreases rapidly with $\tau$ and converges to zero 
after passing through a minimum as $\tau $ approaches infinity. Figure 2
shows the correlation function as a function of $\tau$ for three 
values of temperature and for $\hbar\omega_c=20.0 $ MeV . It 
exhibits a similar behavior as in figure 1. We define the memory time 
$\tau^{(+)}_c$ as the time interval 
during which the correlation function becomes $e^{-1}$ of its value at 
$\tau=0$. The horizontal dashed line in Figs.1 and 2 denotes the position of 
$e^{-1}$.
Figure 3 shows the deduced memory time of the correlation 
function as a function of the cut-off frequency for three 
temperatures studied in Fig.2. 
The memory times for T=0.1 MeV and 1.0 MeV are indistinguishable in the 
figure.  
Figure 4 shows the memory time as a function of the 
temperature for three values of the cut-off frequency. These 
results show that the memory time is nearly independent of temperature 
T and the cut-off frequency $\omega_c$ provided $\hbar\omega_c>5 $ 
MeV, and its magnitude is around  
$\tau_c^{(+)} \approx (0.5-3)*10^{-22}$ second. 
Consequently, when the temperature is comparable or even lower than the 
energy of the quantum zero point motion, a sufficiently large cut-off 
frequency guarantees a short memory time even at such 
low temperatures.

\section{Fluctuation-dissipation theorem for a flat potential or a 
potential well in the short memory time approximation}

We are now ready to derive the fluctuation-dissipation theorem 
in the case, where $C\geq 0$, i.e. for the diffusion process 
in a flat potential or in a potential well. We first change the integration 
variable in eq.(\ref{ROmoment}) from $t_1$ to $\tau=t-t_1$
\begin{eqnarray}
\chi^{(-O)}_{\alpha \alpha}(t)&=&\int^{t-t_0}_0d\tau {\cal S}(t;\tau)
\chi^{(-)}_{\alpha \alpha}(t;\tau)
\label{ROmomentb} 
\end{eqnarray}
where the quantity ${\cal S}(t;\tau)$ is given by,
\begin{eqnarray}
{\cal S}(t,t_1)={\cal S}(t;\tau=t-t_1)=
\frac{1}{{\sqrt {MC}}}sin [\Omega \tau]
\label{mfuncCPb}
\end{eqnarray}
We assume that the memory time of the response function $\tau^{(-)}_c$ 
is much shorter than the relaxation time of the collective motion $\tau_R$.
This condition is satisfied in the weak coupling limit 
if the collective motion 
couples with many incoherent environmental degrees of freedom 
as mentioned before, or if the cut off frequency is high 
in the case of Caldeira-Leggett model as 
shown explicitly by eq.(\ref{clresgd}). 
Then, replacing the upper limit of the integration 
in eq.(\ref{ROmomentb}) by $\infty$, the odd moment of the response
function becomes, 
\begin{eqnarray}
\chi^{(-O)}_{\alpha \alpha}(t)
&\approx & \frac{1}{{\sqrt {MC}}} \int^{\infty}_0d\tau sin \Omega \tau
\chi^{(-)}_{\alpha \alpha}(t;\tau)
\label{ROmomentc} \\
&=& \frac{1}{{\sqrt {MC}}} {\tilde {\chi}}^{(-)}_{\alpha,\alpha}(t;\Omega)
\label{ROmomentd} \\
&=& \frac{1}{{\sqrt {MC}}} \frac{1}{2\hbar}
\{1-e^{-\beta(t)\hbar \Omega}\}
J_{\alpha,\alpha}(t;\Omega) 
\label{ROmomente} 
\end{eqnarray}

Assuming that the memory time of the correlation function is also 
much shorter than the relaxation time of the collective motion, 
i.e. by assuming $\tau_c^{(+)}<<\tau_R$,  
we obtain 
\begin{eqnarray}
\chi^{(+E)}_{\alpha \alpha}(t)
=\frac{1}{2}\frac{1+e^{-\beta(t)\hbar \Omega}}{2}
J_{\alpha,\alpha}(t;\Omega)
\label{CEmomente} 
\end{eqnarray}
Eqs.(\ref{ROmomente}) and (\ref{CEmomente}) 
lead to the well known quantum fluctuation-Dissipation 
relation,  
\begin{eqnarray}
\frac{\chi^{(+E)}_{\alpha \alpha}(t)}{\chi^{(-O)}_{\alpha \alpha}(t)}
=MT^*
\label{FDTP} 
\end{eqnarray}
with the effective temperature given by 
\begin{eqnarray}
T^*&=&\frac{1}{2}\hbar \Omega \cdot coth(\frac{1}{2}\beta(t)\hbar \Omega)
\label{ETa} \\
&=&\left\{
\begin{array}{ll}
T & (T>>\hbar \Omega) \\
\frac{1}{2}\hbar \Omega  & (T<<\hbar \Omega)
\end{array}\right.
\label{ETb} 
\end{eqnarray}

\section{Dissipation coefficient for the Caldeira-Leggett model with a cut-off 
frequency}

In this section, we derive explicit expressions of the 
odd moment of the response function 
for the case of Ohmic dissipation of the Caldeira-Leggett model.

\subsection{Collective motion in a flat potential or in a potential well}

We first consider a collective motion in a flat potential or in a potential 
well. Using the definition of the odd moment of the response function 
given by eq.(\ref{ROmoment}) 
and the expressions derived in sect.V for the response 
function, we obtain for the Ohmic disspation
\begin{eqnarray}
\chi^{(-O)}_{\alpha \alpha}(t)
=\frac{1}{M}(\frac{df}{dq})^2\cdot \eta \cdot g(\Omega)
\label{fricP} 
\end{eqnarray}
Note that the magnitude of the cut-off function $g(\Omega)$
is almost one as long as the cut-off frequency is much 
larger than the curvature parameter $\Omega $ of the 
bare potential of the collective motion. 
On the other hand, the corresponding 
$\chi^{(+E)}_{\alpha \alpha}(t)$ is given by
\begin{eqnarray}
\chi^{(+E)}_{\alpha \alpha}(t)
=\eta \cdot (\frac{df}{dq})^2 \cdot \frac{1}{2} \hbar \Omega 
\cdot g(\Omega) \cdot coth(\frac{1}{2}\beta(t)\hbar \Omega)
\label{emcorre} 
\end{eqnarray}
Eqs.(\ref{fricP}) and (\ref{emcorre}) guarantee the 
fluctuation-dissipation theorem eq.(\ref{FDTP}).

\subsection{Collective motion along a potential barrier}

We next consider a collective motion along a potential barrier, 
where the curvature parameter C is negative. In this case, 
eqs.(\ref{ROmoment}), (\ref{mfuncSN}) and (\ref{clres}) lead to
\begin{eqnarray}
\chi^{(-O)}_{\alpha \alpha}(t)
=\frac{2}{\pi} \eta \frac{1}{\sqrt {M\vert C \vert}}
(\frac{df}{dq})^2 
\int^{t-t_0}_0d\tau sinh(\Omega \tau)
\int^{\infty}_0d\omega sin \omega \tau \cdot \omega \cdot g(\omega)
\label{fricNc} 
\end{eqnarray}
for the case of Ohmic dissipation. For the Gaussian cut-off and the 
Drude reguralization, 
it becomes, 
\begin{eqnarray}
\chi^{(-O)}_{\alpha \alpha}(t)
&=&
\left\{
\begin{array}{ll}
\frac{1}{2{\sqrt {\pi}}} \eta \frac{1}{\sqrt {M\vert C \vert}}
(\frac{df}{dq})^2 \omega^3_c
\int^{t-t_0}_0d\tau sinh(\Omega \tau)\cdot \tau \cdot e^{-\frac{1}{4}
(\omega_c \tau)^2}
 & (Gaussian~~ cut-off) 
\label{fricNd} \\
\eta \frac{1}{\sqrt {M\vert C \vert}}
(\frac{df}{dq})^2 \omega^2_c
\int^{t-t_0}_0d\tau sinh(\Omega \tau) \cdot e^{-\omega_c \tau}
& (Drude~~regularization)
\label{fricNd2} 
\end{array}\right.
\end{eqnarray}
Furthermore, by 
replacing the upper limit of the $\tau $ integral with
$\infty$, we obtain
\begin{eqnarray}
\chi^{(-O)}_{\alpha \alpha}(t)
&=&
\left\{
\begin{array}{ll}
\frac{1}{M}(\frac{df}{dq})^2\cdot \eta \cdot e^{(\frac{\Omega}{\omega_c})^2}
 & (Gaussian~~ cut-off) 
\label{fricNe} \\
\frac{1}{M}(\frac{df}{dq})^2\cdot \eta \cdot \frac{\omega^2_c}
{(\omega_c-\Omega)(\omega_c+\Omega)}
& (Drude~~regularization)
\label{fricNe2} 
\end{array}\right.
\end{eqnarray}
where  $\omega_c>\Omega$ has been assumed for the case of 
the Drude regularization.
These expressions resemble the result for the collective 
motion in a flat potential or in a potential well given by eq.(\ref{fricP}).
It is, however, interesting to notice that the effect of 
the frequency cut-ff appears in a slightly different way.  

The results in this section seem to indicate that the dissipation 
coefficient is almost independent of the temperature of the 
environmental space. However, 
this should not be taken as a general conclusion, but rather as a 
feature of the linear harmonic oscillator coupling model. 
For example, the friction coefficient, which is calculated in a 
random matrix model in \cite{dic} for describing energy dissipation
in deep inelastic heavy-ion collisions, rapidly increases with
temperature. 

\section{Connection between the moments of the response and correlation 
functions for the diffusion along a potential barrier}

Similar expression to eq.(\ref{fricNc}) can be obtained for the 
moment of the correlation function by using eqs.(\ref{CEmoment}),
(\ref{mfuncCN}),(\ref{cspec1}) and (\ref{CLSP}).
It reads
\begin{eqnarray}
\chi^{(+E)}_{\alpha \alpha}(t)
&=&\frac{\hbar}{\pi} \eta (\frac{df}{dq})^2 
\int^{t-t_0}_0d\tau ~cosh(\Omega \tau)
\int^{\infty}_0d\omega ~cos \omega \tau \cdot \omega \cdot g(\omega) 
\cdot \frac{1+e^{-\beta(t)\hbar \omega}}{1-e^{-\beta(t)\hbar \omega}}
\label{flucNc} \\
&=&\frac{\hbar}{\pi} \eta (\frac{df}{dq})^2 \frac{1}{2}
\int^{t-t_0}_0d\tau ~cosh(\Omega \tau)
\int^{\infty}_{-\infty}d\omega ~cos \omega \tau \cdot \omega \cdot g(\omega) 
\cdot \frac{1+e^{-\beta(t)\hbar \omega}}{1-e^{-\beta(t)\hbar \omega}}
\label{flucNc2} \\
&=&\frac{\hbar}{\pi} \eta (\frac{df}{dq})^2 \frac{1}{2}
\int^{t-t_0}_0d\tau ~cosh(\Omega \tau)
\int^{\infty}_{-\infty}d\omega ~cos \omega \tau \cdot \omega \cdot g(\omega) 
\cdot coth\{\frac{\beta(t)\hbar \omega}{2}\}
\label{flucNc3} 
\end{eqnarray}
We first perform the $\tau$ integral analytically, then the $\omega$ integral
using Cauchy's residue theorem. The time integral introduces a pole at 
$\omega=\pm i\Omega$, and the last factor introduces series of poles 
associated with the Matsubara frequency $(2\pi/\hbar \beta)\cdot n_i$, 
$n_i$ being an integer. 
There appears another pole at 
$\omega=\pm i\omega_c$ if we use the Drude regularization for the 
high frequency cut-off. 
The resultant formula reads
\begin{eqnarray}
\chi^{(+E)}_{\alpha \alpha}(t)
=\frac{\hbar}{\pi} \eta (\frac{df}{dq})^2 (Y_1+Y_2+Y_3)
\label{flucNd} 
\end{eqnarray}
with 
\begin{eqnarray}
Y_1&=&-\frac{\pi}{4}\omega^2_c\cdot e^{-\omega_c(t-t_0)}\cdot 
cot(\frac{1}{2}\beta(t)\hbar \omega_c)
\{\frac{1}{\omega_c-\Omega}e^{\Omega(t-t_0)}+
\frac{1}{\omega_c+\Omega}e^{-\Omega(t-t_0)}\}
\label{flucNd1} \\
Y_2&=&\frac{\pi}{\hbar \beta(t)}
\sum_{n=1,2,...}e^{-\frac{\pi\cdot 2n}{\hbar \beta(t)}(t-t_0)}
\frac{\pi \cdot 2n}{\hbar \beta(t)}\cdot \frac{\omega^2_c}
{\omega^2_c-(\frac{\pi\cdot 2n}{\hbar \beta(t)})^2} 
\nonumber \\
&~&\{\frac{1}{\frac{\pi\cdot 2n}{\hbar \beta(t)}-\Omega}e^{\Omega(t-t_0)}+
\frac{1}{\frac{\pi\cdot 2n}{\hbar \beta(t)}+\Omega}e^{-\Omega(t-t_0)}
\}
\label{flucNd2} \\
Y_3&=&\pi \cdot \frac{\omega^2_c}{\omega^2_c-\Omega^2} \cdot 
\frac{1}{2}\Omega \cdot 
cot(\frac{1}{2}\beta(t)\hbar \Omega)
\label{flucNd3} 
\end{eqnarray}
Note that the $Y_1$ term should be ignored in the limit of large $\omega_c$.
If $\omega_c>>T>>\Omega$, then keeping only n=1 term we have
\begin{eqnarray}
Y_2\sim \frac{\pi}{\hbar\beta(t)}e^{(\Omega-\frac{2\pi}{\hbar \beta(t)})(t-t_0)}\label{flucNe} 
\end{eqnarray}
and 
\begin{eqnarray}
Y_3\sim \frac{\pi}{\hbar \beta(t)} 
\label{flucNf} 
\end{eqnarray}
If we keep only the $Y_3$ term in this case, asymptotically, i.e. 
as time goes to infinity, we have
\begin{eqnarray}
\chi^{(+E)}_{\alpha \alpha}(t\rightarrow \infty)
\sim \eta (\frac{df}{dq})^2 \frac{1}{\beta(t)}
\label{flucclasym} 
\end{eqnarray}
which together with eq.(\ref{fricNe2}) leads to the classical 
fluctuation-dissipation theorem.  

The even moment of the correlation function $\chi^{(+E)}(t)$ is the 
momentum diffusion coefficient. At low temperatures, a significant 
deviation in the diffusion coefficient from the classical value 
eq.(\ref{flucclasym}) is expected from  quantal effects \cite{ar94}. 
In order to see this explicitly, in figure 5 we illustrate 
the normalized diffusion coefficient $Y(t)=[Y_1(t)+Y_2(t)+Y_3(t)]
\frac{\hbar}{\pi}$ 
as a function of time $t$ for several temperatures for the collective 
frequency $\hbar\Omega=1.0 $ MeV and the 
cut-off frequency  $\hbar\omega_c=20.0$ MeV. The horizontal lines show 
the classical values determined by temperature $Y_{clas}=T$. 
At high temperatures, the normalized diffusion coefficient
monotonically increases with time and 
gradually approaches to its classical value. However, the
behavior of the diffusion coefficient is very different at low 
temperatures. After a sharp rise, it undershoots the classical value
and may even become negative for large times. We note that a similar 
behaviour of the diffusion coefficient has been discussed also in 
\cite{ar94} in connection with non-Markovian effects.

\section{Summary}

We discussed the quantum effects on the transport coefficients 
for a large amplitude collective motion. To that end, 
we derived a Fokker Planck equation with non-Markovian 
transport coefficients based on a time dependent perturbation theory.
We studied both cases, where the collective motion moves in a flat potentail 
or in a potential well, and along a potential barrier. 
We have shown that our formalism naturally leads to the known generalized 
fluctuation-dissipation theorem which combines the thermal and 
quantal fluctuations when the bare potential is a potential well. 
In the case when the collective motion moves along a potential barrier, 
the well known classical behaviour of the moment 
of the correlation function which governs the fluctuation 
is realized only at high temperatues. Using a 
numerical evaluation of the moment of the correlation function, 
we demonstrated that it significantly deviates 
from the classical expectation at low temperatures, 
both with respect to the time dependence and the asymptotic value. 

\bigskip

\acknowledgements

We thank M. Abe, A. Bulgac, P.G. Rheinhart and 
D.M. Brink for useful discussions. 
This work is supported in part by
the Grand-in-Aid for Scientific Research from 
Ministry of Education, Culture, Sports, Science and Technology 
under Grant No. 12047203 and No. 13640253, 
also under the Special Area Research, Contract number 08640380, 
and by the Japan Society for the Promotion 
of Science for Young Scientists under the contract No. 12006231,
and by the US DOE grant No.DE-FG05-89ER40530. 
N.T. thanks Tennessee Technological University for support 
and hospitality during his visits.

\appendix

\section{Derivation of the reduced von Neumann equation 
in the Galilei transformed coordinate system}

In this Appendix, we explain the major steps to derive 
eq.(\ref{DAvonNeuF}) from eqs.(\ref{DAvonNeu}) through (\ref{DCvonNeu}) by 
the second order perturbation theory. We first multiply a parameter 
$\lambda$ to the fluctuating force ${\hat f}$, and use it as a parameter of 
the perturbation theory.

As usual in the perturbation theory, we expand the reduced density 
operators as
\begin{eqnarray}
{\hat D}_A(t)&=&\sum_{n=0}^\infty \lambda^n {\hat {\cal D}}^{(n)}_A(t)
\label{aada}\\
{\hat D}_B(t)&=&\sum_{n=0}^\infty \lambda^n {\hat {\cal D}}^{(n)}_B(t)
\label{aadb}\\
{\hat D}_C(t)&=&\sum_{n=0}^\infty \lambda^n {\hat {\cal D}}^{(n)}_C(t)
\label{aadc}
\end{eqnarray}

To the zeroth order with respect to $\lambda$, we have
\begin{eqnarray}
i\hbar \frac{\partial}{\partial t}{\hat {\cal D}}^{(0)}_A(t)&=&
[\frac{{\hat p}^2}{2M}+\frac{1}{2}C(t){\hat q}^2,{\hat {\cal D}}^{(0)}_A(t)]
\label{aada0eq}\\
i\hbar \frac{\partial}{\partial t}{\hat {\cal D}}^{(0)}_B(t)&=&0
\label{aadb0eq}\\
i\hbar \frac{\partial}{\partial t}{\hat {\cal D}}^{(0)}_C(t)&=&
[\frac{{\hat p}^2}{2M}+\frac{1}{2}C(t){\hat q}^2,{\hat {\cal D}}^{(0)}_C(t)]
\label{aadc0eq}
\end{eqnarray}
whose solutions are
\begin{eqnarray}
{\hat {\cal D}}^{(0)}_A(t)&=&
{\hat U}^{(0)}_A(t){\hat {\cal D}}^{(0)}_A(t_0)[{\hat U}^{(0)}_A(t)]^\dagger
\label{aada0}\\
{\hat {\cal D}}^{(0)}_B(t)&=&{\hat {\cal D}}^{(0)}_B(t_0)
\label{aadb0}\\
{\hat {\cal D}}^{(0)}_C(t)&=&{\hat U}^{(0)}_A(t){\hat {\cal D}}^{(0)}_C(t_0)
[{\hat U}^{(0)}_A(t)]^\dagger=0
\label{aadc0}
\end{eqnarray}
with 
\begin{eqnarray}
{\hat U}^{(0)}_A(t)=T\{exp[\frac{1}{i\hbar}\int_{t_0}^t
(\frac{{\hat p}^2}{2M}+\frac{1}{2}C(t_1){\hat q}^2)dt_1]\}
\label{aaua0}
\end{eqnarray}
where $T$ is the time ordering operator. We have assumed that the two 
subspaces A and B are decoupled at $t=t_0$, 
so that ${\hat {\cal D}}^{(0)}_C(t_0)=0$.

To the 1st order with respect to $\lambda$, we have
\begin{eqnarray}
i\hbar \frac{\partial}{\partial t}{\hat {\cal D}}^{(1)}_A(t)&=&
Tr_B[{\hat q}\cdot {\hat f}(t), {\hat {\cal D}}^{(0)}_C(t)]+
[\frac{{\hat p}^2}{2M}+\frac{1}{2}C(t){\hat q}^2,{\hat {\cal D}}^{(1)}_A(t)]
\label{aada1eq}\\
i\hbar \frac{\partial}{\partial t}{\hat {\cal D}}^{(1)}_B(t)&=&
Tr_A[{\hat q}\cdot {\hat f}(t), {\hat {\cal D}}^{(0)}_C(t)]
\label{aadb1eq}\\
i\hbar \frac{\partial}{\partial t}{\hat {\cal D}}^{(1)}_C(t)&=&
[{\hat q}\cdot {\hat f}(t), {\hat {\cal D}}^{(0)}_A(t)
{\hat {\cal D}}^{(0)}_B(t)]+
[{\hat q}\cdot {\hat f}(t), {\hat {\cal D}}^{(0)}_C(t)]
\nonumber \\
&& -{\hat {\cal D}}^{(0)}_A(t)Tr_A[{\hat q}\cdot {\hat f}(t), 
{\hat {\cal D}}^{(0)}_C(t)]
-{\hat {\cal D}}^{(0)}_B(t)Tr_B[{\hat q}\cdot {\hat f}(t), 
{\hat {\cal D}}^{(0)}_C(t)] 
\nonumber \\
&& +[\frac{{\hat p}^2}{2M}+\frac{1}{2}C(t){\hat q}^2,
{\hat {\cal D}}^{(1)}_C(t)],
\label{aadc1eq}
\end{eqnarray}
where eq.(\ref{aadc1eq}) can be solved by the method of variational constant 
by putting
\begin{eqnarray}
{\hat {\cal D}}^{(1)}_C(t)={\hat U}^{(0)}_A(t){\hat C}(t)
[{\hat U}^{(0)}_A(t)]^\dagger
\label{aadc1vc}
\end{eqnarray}
The result reads
\begin{eqnarray}
{\hat {\cal D}}^{(1)}_C(t)
&=&{\hat U}^{(0)}_A(t){\hat {\cal D}}^{(1)}_C(t_0)
[{\hat U}^{(0)}_A(t)]^\dagger
\nonumber \\
&+&\frac{1}{i\hbar}\int_{t_0}^t ~dt_1
[{\hat U}^{(0)}_A(t,t_1){\hat q}[{\hat U}^{(0)}_A(t,t_1)]^\dagger
\cdot {\hat f}(t_1), 
{\hat U}^{(0)}_A(t,t_1)
{\hat {\cal D}}^{(0)}_A(t_1)
[{\hat U}^{(0)}_A(t,t_1)]^\dagger{\hat {\cal D}}^{(0)}_B(t_1)]
\label{aadc1sol}
\end{eqnarray}
with 
\begin{eqnarray}
{\hat U}^{(0)}_A(t,t_1)={\hat U}^{(0)}_A(t){\hat U}^{(0)}_A(t_1)^\dagger
\label{aatevop}
\end{eqnarray}

Now we note that we have 
\begin{eqnarray}
i\hbar \frac{\partial}{\partial t}{\hat D}^{(2)}_A(t)=
Tr_B[{\hat q}\cdot {\hat f}(t), {\hat D}^{(1)}_C(t)]+
[\frac{{\hat p}^2}{2M}+\frac{1}{2}C(t){\hat q}^2,{\hat D}^{(2)}_A(t)]
\label{aasoeq}
\end{eqnarray}
in the second order approximation. Note that  
\begin{eqnarray}
{\hat D}^{(2)}_A(t)&=&{\hat {\cal D}}^{(0)}_A(t)+
\lambda {\hat {\cal D}}^{(1)}_A(t)+\lambda^2 {\hat {\cal D}}^{(2)}_A(t)
\label{aadrel1}\\
{\hat D}^{(1)}_C(t)&=&{\hat {\cal D}}^{(0)}_C(t)+
\lambda {\hat {\cal D}}^{(1)}_C(t)
\label{aadrel2}
\end{eqnarray}
Eq.(\ref{aadc1sol}) then leads to
\begin{eqnarray}
i\hbar \frac{\partial}{\partial t}{\hat {\cal D}}^{(2)}_A(t)&=&
\frac{1}{i\hbar}\int_{t_0}^t~dt_1 
Tr_B[{\hat q}\cdot{\hat f}(t), 
[{\hat U}^{(0)}_A(t,t_1){\hat q}[{\hat U}^{(0)}_A(t,t_1)]^\dagger
\cdot {\hat f}(t_1),
\nonumber \\
&&{\hat U}^{(0)}_A(t,t_1){\hat {\cal D}}^{(0)}_A(t_1)
[{\hat U}^{(0)}_A(t,t_1)]^\dagger {\hat {\cal D}}^{(0)}_B(t_1)]]
+[\frac{{\hat p}^2}{2M}+\frac{1}{2}C(t){\hat q}^2,{\hat D}^{(2)}_A(t)]
\label{aada2eq1}\\
&\approx& \frac{1}{i\hbar}\int_{t_0}^t~dt_1 
Tr_B[{\hat q}\cdot{\hat f}(t), 
[{\hat U}^{(0)}_A(t,t_1){\hat q}[{\hat U}^{(0)}_A(t,t_1)]^\dagger
\cdot {\hat f}(t_1),{\hat D}^{(2)}_A(t){\hat D}^{(2)}_B(t_1)]]
\nonumber \\
&+&[\frac{{\hat p}^2}{2M}+\frac{1}{2}C(t){\hat q}^2,{\hat D}^{(2)}_A(t)]
\label{aada2eq2}
\end{eqnarray}
The transformation from eq.(\ref{aada2eq1}) to eq.(\ref{aada2eq2}) 
is consistent with the second order perturbation theory.
It is now straightforward to transform eq.(\ref{aada2eq2}) into the form of 
eq.(\ref{DAvonNeuF}).

\section{Symmetry property of the spectral function}

The spectral function is given by 
\begin{eqnarray}
J_{\alpha,\beta}(t;\omega)
=\int^{+\infty}_{-\infty}d\tau e^{i\omega \tau}
Tr_B({\hat F}_\alpha(t)e^{-\frac{i}{\hbar}{\hat h}_B(t)\tau}{\hat F}_\beta(t)
e^{\frac{i}{\hbar}{\hat h}_B(t)\tau}{\hat \rho}_B(t))
\label{aa1} 
\end{eqnarray}
for a slow collective motion. 
We assume that the density operator of the bath space ${\hat \rho}_B(t)$ 
can be approximated by a canonical distribution 
$e^{\beta(t)({\cal F}-{\hat h}_B(t))}$. The spectral function then reads
\begin{eqnarray}
J_{\alpha,\beta}(t;\omega)
=\int^{+\infty}_{-\infty}d\tau e^{i\omega \tau}
Tr_B({\hat F}_\alpha(t)e^{-\frac{i}{\hbar}{\hat h}_B(t)\tau}{\hat F}_\beta(t)
e^{\frac{i}{\hbar}{\hat h}_B(t)\tau}e^{\beta(t)({\cal F}-{\hat h}_B(t))})
\label{aa2} 
\end{eqnarray}
The key step is to insert two delta functions as follows
\begin{eqnarray}
J_{\alpha,\beta}(t;\omega)
&=&\int^{+\infty}_{-\infty}d\tau e^{i\omega \tau}
Tr_B({\hat F}_\alpha(t)e^{-\frac{i}{\hbar}{\hat h}_B(t)\tau}
\int^{\infty}_{-\infty} dE\delta(E-{\hat h}_B(t)){\hat F}_\beta(t)
\nonumber \\
&~& \int^{\infty}_{-\infty}  dE'\delta(E'-{\hat h}_B(t))
e^{\frac{i}{\hbar}{\hat h}_B(t)\tau}e^{\beta(t)({\cal F}-{\hat h}_B(t))})
\label{aa3} 
\end{eqnarray}
We then obtain
\begin{eqnarray}
J_{\alpha,\beta}(t;\omega)
=2\pi i \int^{+\infty}_{-\infty}dE \int^{+\infty}_{-\infty}dE'
\delta(E-E'-\hbar \omega) j_{\alpha,\beta}(E,E')\rho_{eq}(E')
\label{aa4} 
\end{eqnarray}
where 
\begin{eqnarray}
& &j_{\alpha,\beta}(E,E')=
Tr_B({\hat F}_\alpha(t)\delta(E-{\hat h}_B(t)){\hat F}_\beta(t)
\delta(E'-{\hat h}_B(t))
\label{aa5} \\
& &\rho_{eq}(E)=e^{\beta(t)({\cal F}-E)}
\label{aa6} 
\end{eqnarray}

One can easily prove the following symmetry relations
\begin{eqnarray}
& &j_{\beta,\alpha}(E',E)=j_{\alpha,\beta}(E,E')
\label{aa7} \\
& &[j_{\alpha,\beta}(E,E')]^*
=j_{F^\dagger_\alpha,F^\dagger_\beta}(E',E)
\label{aa8} 
\end{eqnarray}
We now consider
\begin{eqnarray}
J_{\beta,\alpha}(t;\omega)
=2\pi i \int^{+\infty}_{-\infty}dE \int^{+\infty}_{-\infty}dE'
\delta(E-E'-\hbar \omega) j_{\beta,\alpha}(E,E')\rho_{eq}(E')
\label{aa9} 
\end{eqnarray}
Using eqs.(\ref{aa7}) and (\ref{aa6}) and interchanging the integration 
variables $E$ and $E'$, one can prove that
\begin{eqnarray}
J_{\beta,\alpha}(t;\omega)
=e^{\beta(t)\hbar \omega}J_{\alpha,\beta}(t;-\omega)
\label{aa10} 
\end{eqnarray}
This proves eq.(\ref{symspec}) in the main text.

\section{Basic formulae for the Feynman-Vernon-Caldeira-Leggett model in the 
linear response theory}

In this Appendix, we collect several basic formulae 
for the case, where the environment is represented by an 
aggregate of harmonic oscillators, and where the collective motion 
couples them linearly with respect to their coordinates
\cite{cl,fv}. 
The unperturbed Hamiltonian of the bath space ${\hat H}_B$, 
the coupling Hamiltonian ${\hat V}_c(q(t),{\hat x}$ and the 
force operator ${\hat F}_\alpha(q(t),{\hat x})$  read
\begin{eqnarray}
& &{\hat H}_B=\sum_i(\frac{1}{2m_1}{\hat p}^2_i+
\frac{1}{2}m_i\omega^2_i{\hat x}^2_i)
\label{CLHB} \\
& &{\hat V}_c(q(t),{\hat x})=f(q(t))\sum_ic_i{\hat x}_i
\label{CLVc} \\
& &{\hat F}_\alpha(q(t),{\hat x})=\frac{df(q(t))}{dq_\alpha}
\sum_ic_i\{{\hat x}_i-Tr({\hat x}_i{\hat \rho}_B(t))\}
\label{CLF} 
\end{eqnarray}

In the linear response theory, we 
replace the ${\hat \rho}_B(t)$ and ${\hat h}_B(t)$ on the r.h.s. of 
eq.(\ref{aspecfun}) by the canonical distribution 
$e^{\beta(t)({\cal F}-{\hat H}_B)}$ and ${\hat H}_B$, respectively.
Eqs.(\ref{CLHB}) through (\ref{CLF}) then lead to 
\begin{eqnarray}
J^{(CLLR)}_{\alpha \alpha}(t;\omega)
=2\pi \cdot (\frac{df}{dq})^2\cdot \sum_i c^2_i\frac{\hbar}{2m_i\omega_i}
\{\delta(\omega+\omega_i)\frac{e^{-\beta(t)\hbar \omega_i}}
{1-e^{-\beta(t)\hbar \omega_i}}+
\delta(\omega-\omega_i)\frac{1}
{1-e^{-\beta(t)\hbar \omega_i}}
\}
\label{CLSP}
\end{eqnarray}
One can easily prove that this expression satisfies the 
symmetry property eq.(\ref{symspec}). 
It would be physically reasonable to assume that all $\omega_i$ are 
positive. Then, for $\omega>0$, the relevant part of the 
spectral function is 
\begin{eqnarray}
J^{(CLLR)}_{\alpha \alpha}(t;\omega)
=2\pi \cdot (\frac{df}{dq})^2 \cdot \sum_i c^2_i\frac{\hbar}{2m_i\omega_i}
\delta(\omega-\omega_i)\frac{1}
{1-e^{-\beta(t)\hbar \omega_i}} \hskip 1cm (\omega>0)   
\label{CLSPb}
\end{eqnarray}

We now define the spectral density $J(\omega)$ by
\begin{eqnarray}
J(\omega)=\sum_i c^2_i\frac{1}{m_i\omega_i}\delta(\omega-\omega_i)
\label{spdensity}
\end{eqnarray}
Following \cite{cl}, we assume that
\begin{eqnarray}
J(\omega)=\frac{2}{\pi}\eta 
\frac{1}{\omega^{n-1}_0}
\omega^n g(\omega)
\label{sdcl}
\end{eqnarray}
The cases of n=1 and n=3 correspond to the Ohmic and super Ohmic 
dissipations, respectively. In eq.(\ref{sdcl}),we have introduced 
$\omega_0$ in oredr to keep the dimension of $\eta $ independent of $n$.
$g(\omega)$ is the cut-off function, for which we assume two-types of 
functional forms
\begin{eqnarray}
g(\omega)&=&g_1(\omega)=e^{-(\frac{\omega}{\omega_c})^2}
\label{cutofffg} \\ 
g(\omega)&=&g_2(\omega)=\frac{1}{1+(\frac{\omega}{\omega_c})^2}
\label{cutofffg2}. 
\end{eqnarray}
We call the first and the second the Gaussian cut off and 
the Drude regularization, respectively.

\section{General expression of the m-th moment 
of the response function for the Caldeira-Leggett model} 

We consider 
\begin{eqnarray}
\chi^{(-m)}_n(t)
=\frac{2}{\pi}\eta (\frac{df}{dq})^2
\frac{1}{\omega^{n-1}_0}
\int^{\infty}_0d\tau \tau^m
\int^{\infty}_0d\omega sin \omega \tau 
\cdot \omega^n \cdot g(\omega)
\label{reshm} 
\end{eqnarray}
where both n and m are integers; n,m=1,2,.... 
This will be the m-th moment of the response function if the 
spectral density is given by
\begin{eqnarray}
\sum_i c^2_i\frac{1}{m_i\omega_i}\delta(\omega-\omega_i)=\frac{2}{\pi}\eta 
\frac{1}{\omega^{n-1}_0}
\omega^n g(\omega)
\label{sOhmic}
\end{eqnarray}
One can show
\begin{eqnarray}
\chi^{(-m)}_n(t)
&=&\eta (\frac{df}{dq})^2
\frac{1}{\omega^{n-1}_0}
(-1)^{[\frac{m}{2}]+1}
\int^{\infty}_{-\infty}d\omega \omega^n 
\cdot g(\omega) \cdot \frac{d^m}{d\omega^m}\delta(\omega)
\label{reshm2} \\
&=&\eta (\frac{df}{dq})^2
\frac{1}{\omega^{n-1}_0}
(-1)^{[\frac{m}{2}]+1}(-1)^m
[\frac{d^m}{d\omega^m}\{\omega^n g(\omega)\}]_{\omega=0}
\label{reshm3} 
\end{eqnarray}

Epecially, 
\begin{eqnarray}
\chi^{(-1)}_1(t)&=&\eta (\frac{df}{dq})^2 g(0)
\label{resO1} \\
\chi^{(-1)}_3(t)&=&0
\label{ressO1} 
\end{eqnarray}
Eq.(\ref{resO1}) agrees with the results eq.(\ref{RES1-O}) together with 
eqs.(\ref{fricP}) and (\ref{fricNe}).
On the other hand, eq.(\ref{ressO1}) means 
that there exists no linear friction 
in the case of super Ohmic dissipation. 
Eq.(\ref{reshm3}) shows that the lowest order friction is n-th 
order friction if the spectral density is proportional to $\omega^n$ as 
given in eq.(\ref{sOhmic}) and the friction coefficient 
is given by the n-th moment of the response function.


\newpage
\noindent
Figure captions

\medskip

\noindent
Figure 1. The normalized correlation function $\chi^{(+)}(\tau)$ as a 
function of $\tau $ for three values of the cut-off frequency and 
for a temperature T=1 MeV. The horizontal dashed line shows the 
value $e^{-1}$.

\medskip

\noindent
Figure 2. The normalized correlation function $\chi^{(+)}(\tau)$ as a 
function of $\tau $ for three values of temperature and for a 
cut-off frequency $\hbar\omega_c=20.0 $ MeV. The horizontal dashed 
line shows the value $e^{-1}$.

\medskip

\noindent
Figure 3. Memory time of the correlation function 
$\tau_c^{(+)}$ as a function of cut-off 
frequency $\hbar\omega_c$  for three values of temperature.

\medskip

\noindent
Figure 4. Memory time of the correlation function 
$\tau_c^{(+)}$ as a function of temperature
for three values of the cut-off frequency $\hbar\omega_c$.

\medskip

\noindent
Figure 5. The normalized diffusion coefficent 
$Y(t)$ as a function of 
$t$ for several values of temperature and for a collective frequency 
$\hbar\Omega=1.0 $ MeV and a cut-off frequency  $\hbar\omega_c=20.0$ 
MeV. The horizontal dashed lines show the corresponding classical 
values.


\begin{thebibliography}{9}
\bibitem{weiss} U. Weiss, QUANTUM DISSIPATIVE SYSTEMS (World Scientific,
1993,Singapore).

\bibitem{llsp} L.D. Landau and E.M. Lifshitz, Statistical Physics, 
Part 1, 3rd edition (Butterworth/Heinemann, 1997, Oxford) p. 384

\bibitem{ar94} S. Ayik and J. Randrup, Phys. Rev. C50(1994) 2947. 

\bibitem{hh00} H. Hofmann and F.A. Ivanyuk, Phys. Rev. Letts. 82(1999) 4603.

\bibitem{aygs98} S. Ayik, O. Yilmaz, A. Gokalp and P. Schuck, 
Phys. Rev. C58(1998) 1594. 

\bibitem{Kra40} H.A. Kramers, Physica VII,no.4 (1940) 284.

\bibitem{awo} Y. Abe, et al., AIP conrefernce series Vol.597(2001) p.209. 

\bibitem{tak81} N. Takigawa, K. Niita, Y. Okuhara and S. Yoshida,
Nucl. Phys. A371 (1981), p.130.  

\bibitem{kubo} Statistical Mechanics, eds. M. Toda and R. Kubo 
(Iwanami, Tokyo, 1978). 

\bibitem{cl} A.O. Caldeira and A.J. Leggett, PRL 46(1981)211
;Ann. Phys. (N.Y.) 149(1983) 374.  


\bibitem{Bec78} R. Becker, Theorie der W${\ddot a}$rme (Springer-Verlag, 
Berlin,1978) p.288

\bibitem{van Hove55} L. van Hove, Physica 21(1955) 517.

\bibitem{acw} D. Agassi, C.M. Co and H. Weidenm\"uller, Ann. Phys.
(N.Y.) 107(1977)140;117(1979)404.

\bibitem{fv} R.P. Feynmann and F.L. Vernon, Ann. Phys. (NY) 24(1963)118;
R.P. Feynman ``Statistical Mechanics"(Benjamin, Reading, Mass.,1972);
R.P. Feynman and A.R. Hibbs, ``Quantum Mechanics and Path Integrals"
(Mc Graw-Hill, New York,1965)
  


\bibitem{dic} K. Niita and N. Takigawa, Nucl. Phys. A397 (1983), p.141-160.

\end{thebibliography}
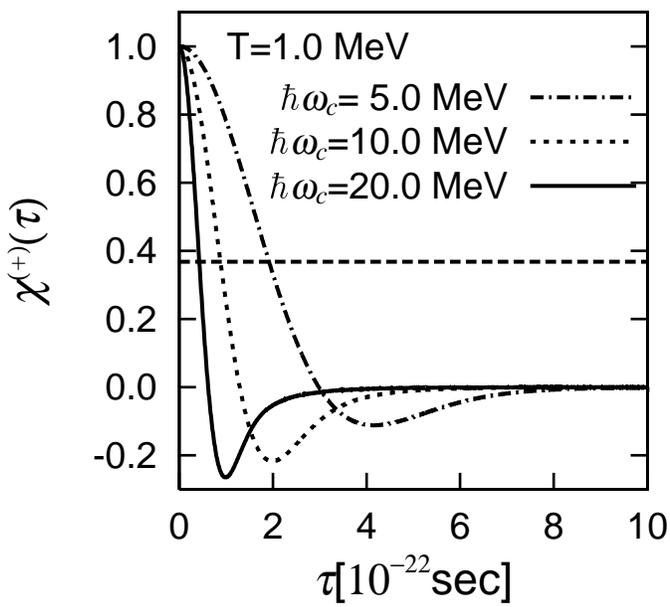
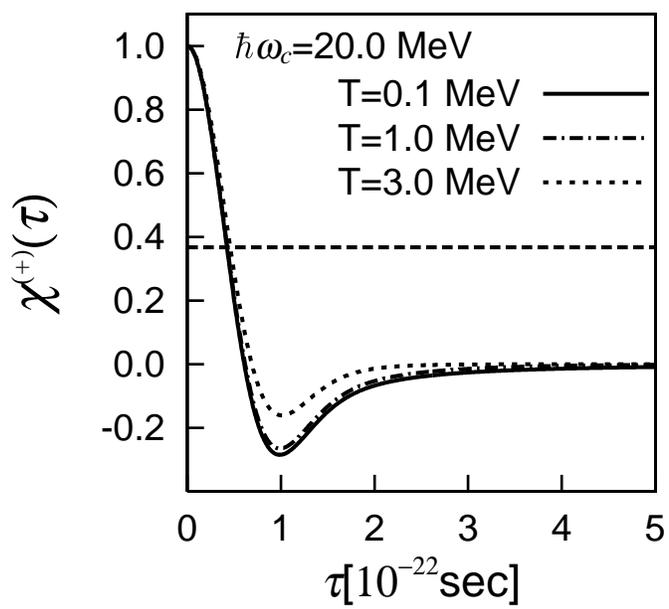
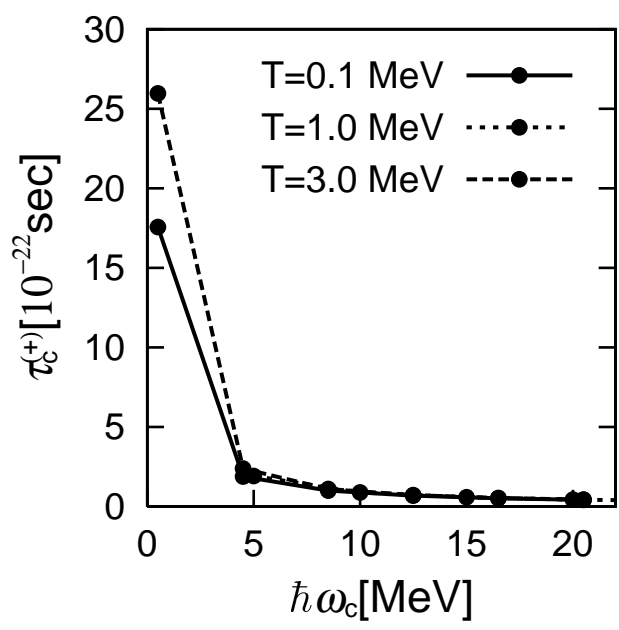
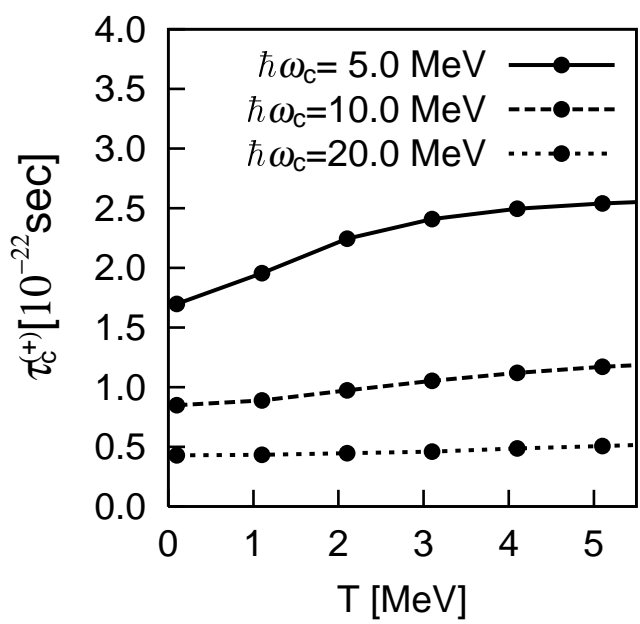
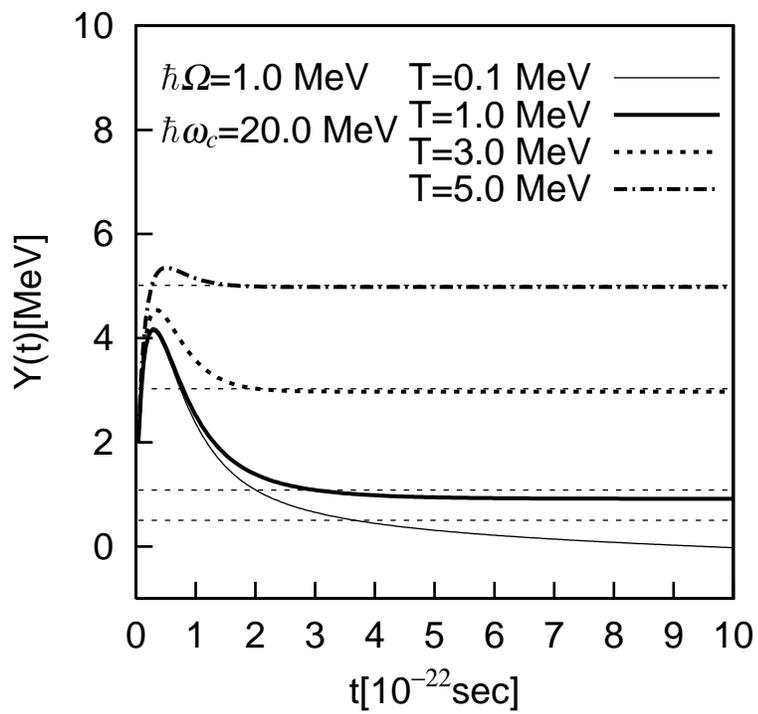
\end{document}